\documentclass[aps,pre,reprint,groupedaddress,floatfix]{revtex4-2}
\usepackage[colorlinks=true,allcolors=blue]{hyperref} 
\usepackage{graphicx} 
\usepackage{bm}
\usepackage{amsmath, amssymb}
\usepackage{braket}
\usepackage{appendix}

\newcommand{\Figref}[1]{Fig.~\ref{#1}}

\newcommand{\Eqref}[1]{Eq.~(\ref{#1})}
\newcommand{\Eqsref}[1]{Eqs.~(\ref{#1})}

\begin{document}
\title{Steady Motions of Single Spherical Microswimmers in Non-Newtonian Fluids}
\author{Takuya Kobayashi}
\email{kobayashi@cheme.kyoto-u.ac.jp}
\author{Ryoichi Yamamoto}
\email{ryoichi@cheme.kyoto-u.ac.jp}
\affiliation{
Department of Chemical Engineering, Kyoto University, Kyoto 615-8510, Japan
}

\date{\today}

\begin{abstract}
In biological systems, microswimmers often propel themselves through complex media. However, many aspects of swimming mechanisms in non-Newtonian fluids remain unclear. This study considers the propulsion of two types of single spherical microswimmers (squirmers) in shear-thickening and shear-thinning fluids. The slip-driven squirmer propels faster/slower in shear-thickening/thinning fluids than in Newtonian fluids [C. Datt {\it et al.}, ``Squirming through shear-thinning fluids,'' J. Fluid Mech. \textbf{784}, R1 (2015)]. In contrast, we discovered that a traction-driven squirmer exhibits the opposite trend, moving slower/faster in shear-thickening/thinning fluids than in Newtonian fluids.
In addition, we have shown theoretically that Purcell's scallop theorem does not hold in non-Newtonian fluids when a squirmer with reciprocal surface motions is used. The present findings open up possibilities for the design of new types of microswimmers that can achieve translational motion from a single reciprocal motion in non-Newtonian fluids.
\end{abstract}

\maketitle

\section{Introduction}
    Microswimmers exemplify soft active matter~\cite{Marchetti2013-yq} and have recently gained attention for their essential roles in various biological processes, including female reproduction~\cite{Leftwich2024}, and their potential for biomedical applications, such as microcargo transport, targeted drug delivery, artificial insemination, and microsurgery~\cite{Bunea2020-fv}. 
    In microscale swimming, viscous forces are much more significant than inertial forces are, placing the systems in the low Reynolds number regime. While we know much about how microswimmers move in simple Newtonian fluids~\cite{Lauga2009-jt}, an understanding of their movement in complex fluids is still lacking. Many fluids in the human body, such as mucus~\cite{Lai2009-oz}, vitreous humor in the eye~\cite{Silva2017-ss} and blood~\cite{Lynch2022-yd}, exhibit non-Newtonian properties. Natural microswimmers, such as bacteria~\cite{Ali2016-dh} and sperm cells~\cite{Fauci2006-dx}, often encounter non-Newtonian rheological behaviors. 
    These complex fluids differ from simple Newtonian fluids because of their rheological properties, such as normal stress differences and shear-rate-dependent viscosity. Normal stress differences, arising from fluid elasticity, can affect the speed of microswimmers. Typically, microswimmers, such as squirmers~\cite{Zhu2011-eo, Zhu2012-wp} and swimming sheet and filament models~\cite{Lauga2007-pi, Fu2007-ks, Fu2009-qj}, tend to move slower in non-Newtonian fluids than in Newtonian fluids. However, they can achieve higher speeds by coupling their chirality and normal stress differences~\cite{Binagia2020-dn, Housiadas2021-qe, Kobayashi2023-ad, Kobayashi2024-tn}. 
    The shear rate-dependent viscosity can also affect the swimming speed, depending on the specific type of swimmer~\cite{Montenegro-Johnson2013-uj, Datt2015-yc, Nganguia2020-ss}. Understanding microswimmers' locomotion in complex fluids is crucial for designing artificial microrobots for biomedical applications such as targeted drug delivery~\cite{Nelson2010-uz}.

    For spherical microswimmers, many previous theoretical and numerical investigations have focused on slip-driven squirmers~\cite{Lighthill1952-ui, Blake1971-ws, Pedley2016-wt, Ishikawa2024-vp}. However, traction-driven squirmers are more realistic when modeling organisms such as ciliates~\cite{Ishikawa2020-dl, Ishikawa2024-vp}. To the best of our knowledge, much less research has been conducted on traction-driven microswimmers, with only a few studies examining microswimming in Stokes fluids~\cite{Ishikawa2020-dl, Nasouri2021-tf, Daddi-Moussa-Ider2023-tv} and Stokes fluids with odd viscosities~\cite{Hosaka2023-tt}. The locomotion problems for traction-driven microswimmers in complex fluids, which exhibit normal stress differences or shear-thickening/thinning effects, remain unclear, leaving these problems as ongoing challenges.

    Purcell's scallop theorem states that microswimmers with a single degree of freedom in their deformation cannot achieve net locomotion in Stokes fluids~\cite{Purcell1977-ab}.
    Ishimoto {\it et al.} provided a coordinate-based proof of this theorem, including body rotation~\cite{Ishimoto2012-iv}. However, most biomedical fluids are non-Newtonian, and the scallop theorem does not apply.
    In viscoelastic fluids, microswimmers with a single reciprocal motion can achieve net locomotion. The snowman model, which consists of two different-sized rotating spheres, can move due to normal stress differences~\cite{Pak2012-rk, Puente-Velazquez2019-fs, Binagia2021-fi, Kroo2022-qm}. Additionally, Lauga analytically demonstrated that normal stress differences can break down the scallop theorem for a squirmer with time-reversible deformation~\cite{Lauga2009-df}.
    Previous experimental investigations have shown that the shear-thinning effect can also break down the scallop theorem, allowing a single-hinged scallop to swim in both shear-thickening and shear-thinning fluids~\cite{Qiu2014-nm, Han2020-hy}. 

    To better understand the propulsion of a single microswimmer in non-Newtonian fluids with shear-rate-dependent viscosity, we studied two types of surface-driven microswimmers: (i) a slip-driven squirmer and (ii) a traction-driven squirmer. Using the reciprocal theorem and asymptotic analysis, we calculated the swimming speed of these microswimmers in fluids described by the Carreau--Yasuda model. Our findings show that traction-driven squirmers move slower/faster, whereas slip-driven squirmers move faster/slower in shear-thickening/thinning fluids than in Newtonian fluids~\cite{Datt2015-yc}. We also analyzed the differences in swimming speed between slip- and traction-driven squirmers by decomposing the problem into drag and thrust forces.
    Additionally, we investigated whether the scallop theorem applies to these microswimmers in shear-thickening/thinning fluids and found that it does not apply to either type. 

\section{
Lorentz' Reciprocal Theorem for complex fluids
}
\begin{figure}[tb]
        \centering
        \includegraphics[width=\linewidth]{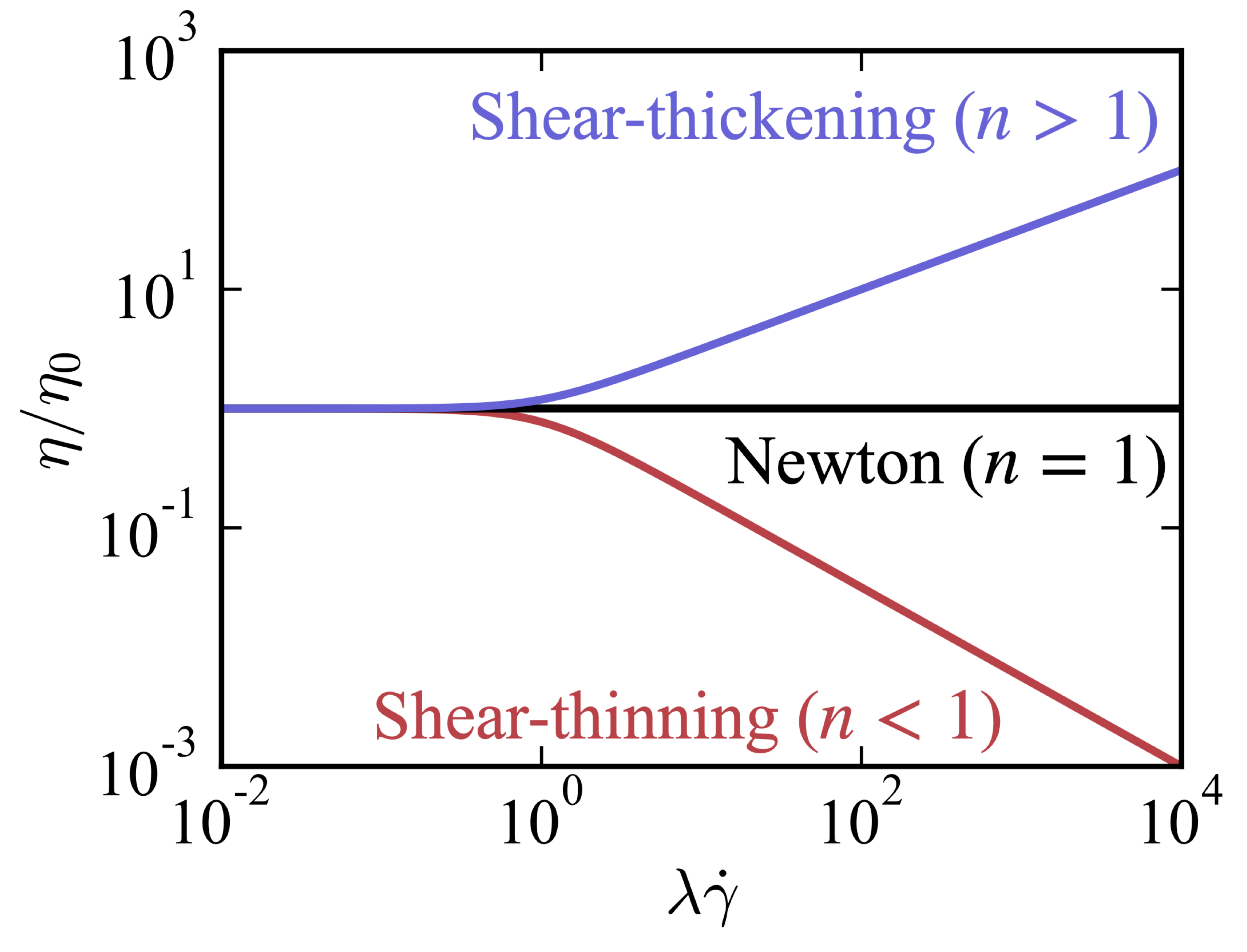}
        \caption{Viscosity $\eta$ is governed by the Carreau--Yasuda law with an infinite-shear-rate viscosity $\eta_\infty = 0$.}
        \label{fig:Carreau-Yasuda}
    \end{figure}

    Here, we revisit integral theorems for locomotion in non-Newtonian fluids~\cite{Lauga2014-tg, Elfring2015, Elfring2017-zf}. 
    In this study, we consider incompressible viscous fluids with shear-rate-dependent viscosity. Since fluid inertia can be ignored in the microhydrodynamic regime, the governing equations for the fluids are as follows:
    \begin{subequations}
        \begin{gather}
            \bm{\nabla}\cdot\bm{u} = 0,\\
            \bm{\nabla}\cdot\bm{\sigma} = \bm{0},\\
            \bm{\sigma} = -p\bm{1} + \eta\dot{\bm{\gamma}},
        \end{gather}
    \end{subequations}
    where $p$ and $\dot{\bm{\gamma}} = \bm{\nabla}\bm{u} + (\bm{\nabla}\bm{u})^T$ are the pressure and strain rate, respectively. 
    To account for the viscosity, which depends on the shear rate, i.e., shear-thickening and shear-thinning rheological properties, we use the Carreau--Yasuda constitutive law~\cite{Bird1987-qc}, described by
    \begin{align}
        \eta = \eta_\infty + (\eta_0 - \eta_\infty)\left(1 + \lambda^2|\dot{\gamma}|^2\right)^{(n - 1) / 2},
    \end{align}
    where $\eta_0$ and $\eta_\infty$ represent the viscosity at zero and infinite shear rates, respectively. The magnitude of the strain rate tensor is given by $|\dot{\gamma}|^2 = \dot{\gamma}_{ij}\dot{\gamma}_{ij} / 2$. The power law index $n$ indicates the degree of shear thickening $(n > 1)$ and thinning $(n < 1)$. The relaxation time $\lambda$ determines the strain rate at which non-Newtonian behavior becomes significant (see \Figref{fig:Carreau-Yasuda}). Experimental data on human cervical mucus can be effectively modeled via the Carreau--Yasuda model~\cite{Hwang1969, Velez-Cordero2013-pi}.

    To derive the integral theorem, we choose $\hat{\bm{u}}$ and $\hat{\bm{\sigma}}$, governed by the Stokes equation with constant viscosity $\eta_0$, as the auxiliary problem.
    We start with the identities,
    \begin{subequations}
        \begin{gather}
            \bm{\nabla}\cdot(\bm{\sigma}\cdot\hat{\bm{u}}) = (\bm{\nabla}\cdot\bm{\sigma})\cdot\hat{\bm{u}} + \bm{\sigma}:\bm{\nabla}\hat{\bm{u}} = \bm{\sigma}:\bm{\nabla}\hat{\bm{u}},\\
            \bm{\nabla}\cdot(\hat{\bm{\sigma}}\cdot\bm{u}) = (\bm{\nabla}\cdot\hat{\bm{\sigma}})\cdot\bm{u} + \hat{\bm{\sigma}}:\bm{\nabla}\bm{u} = \hat{\bm{\sigma}}:\bm{\nabla}\bm{u},
        \end{gather}
    \end{subequations}
    where $\bm{A}:\bm{B} = A_{ij}B_{ij}$, and we use $\bm{\nabla}\cdot\bm{\sigma} = \bm{\nabla}\cdot\hat{\bm{\sigma}} = \bm{0}$.  
    We then obtain the identity
    \begin{align}
        \bm{\nabla}\cdot(\bm{\sigma}\cdot\hat{\bm{u}}) - \bm{\nabla}\cdot(\hat{\bm{\sigma}}\cdot\bm{u}) = \bm{\sigma}:\bm{\nabla}\hat{\bm{u}} - \hat{\bm{\sigma}}:\bm{\nabla}\bm{u}.
    \end{align}
    By integrating this identity over the fluid volume ${V}$ and using the divergence theorem, we derive the integral theorem for complex fluids:
    \begin{align}\label{eq:integral_theorem}
        \begin{split}
            -\int_{S}\bm{n}\cdot\bm{\sigma}\cdot\hat{\bm{u}}\ dS + \int_{S}\bm{n}\cdot\hat{\bm{\sigma}}\cdot\bm{u}\ dS \\
            = \int_{V}(\bm{\sigma}:\bm{\nabla}\hat{\bm{u}} - \hat{\bm{\sigma}}:\bm{\nabla}\bm{u})\ dV,
        \end{split}
    \end{align}
    where ${S}$ is the surface of a microswimmer, and $\bm{n}$ is the normal surface vector pointing into the fluid.
    Using a perturbation expansion for small Carreau numbers ${\rm Cu} = \lambda |\dot{\gamma}_0|$, we can express the right-hand side of \Eqref{eq:integral_theorem} up to ${O}({\rm Cu}^2)$,
    \begin{align}\label{eq:perturbation}
        \bm{\sigma}:\bm{\nabla}\hat{\bm{u}} - \hat{\bm{\sigma}}:\bm{\nabla}\bm{u} = \underbrace{\frac{(\eta_0 - \eta_\infty)(n - 1)}{2}{\rm Cu}^2 \dot{\bm{\gamma}}_0}_{\bm{\Sigma}}:\bm{\nabla}\hat{\bm{u}},
    \end{align}
    where $\dot{\bm{\gamma}}_0$ is the strain-rate tensor corresponding to the velocity field in the Stokes limit, and $\bm{\Sigma}$ represents the non-Newtonian contribution.
    From \Eqsref{eq:integral_theorem} and~\eqref{eq:perturbation}, we derive the integral theorem for complex fluids, followed by the Carreau--Yasuda constitutive law,
    \begin{align}\label{eq:theorem}
        -\int_{S}\bm{n}\cdot\bm{\sigma}\cdot\hat{\bm{u}}\ dS + \int_{S}\bm{n}\cdot\hat{\bm{\sigma}}\cdot\bm{u}\ dS 
        = \int_{V} \bm{\Sigma}:\bm{\nabla}\hat{\bm{u}}\ dV.
    \end{align}
    This integral theorem describes the behavior of complex fluids governed by the Carreau--Yasuda model, taking into account the small Carreau number approximation.

\section{Surface-driven microswimmer}
    \begin{figure}[tb]
        \centering
        \includegraphics[width=\linewidth]{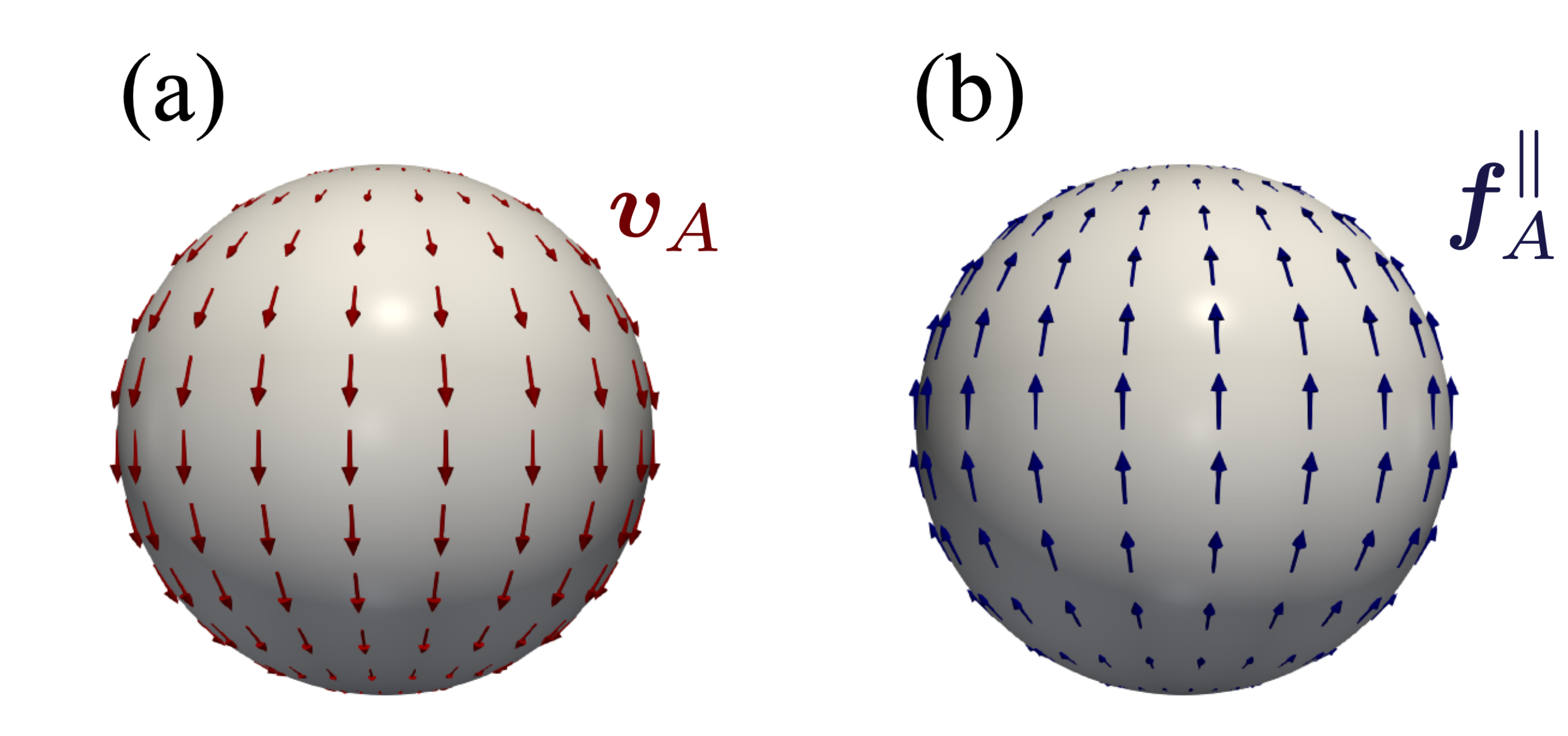}
        \caption{Schematic illustrations of (a) a slip-driven squirmer and (b) a traction-driven squirmer. (a) $\bm{v}_A = -v_0(\bm{1} - \bm{nn})\cdot\mathbf{p}$, and (b) $\bm{f}^\parallel_A = f_0(\bm{1} - \bm{nn})\cdot\mathbf{p}$.}
        \label{fig:swimmer}
    \end{figure}
    We employ the integral theorem \Eqref{eq:theorem} to determine the velocity of two types of microswimmers (slip-driven and traction-driven squirmers, as shown in~\Figref{fig:swimmer}) in viscous fluids with shear-rate-dependent viscosities. 
    We consider a microswimmer with axisymmetric surface squirming motion, which results in axisymmetric surface slip velocity for the slip-driven squirmer and force density for the traction-driven squirmer. This ensures that both microswimmers have only translational motion ($\bm{\Omega} = \bm{0}$), as previously assumed~\cite{Datt2015-yc}.
    We choose two problems: (i) for a self-propelled spherical swimmer with velocity $\bm{U}$ in shear-thickening and shear-thinning fluids and (ii) for a passive body of the same shape dragged with velocity $\hat{\bm{U}}$ in Stokes fluids.
    The boundary conditions at the surfaces are as follows:
    \begin{subequations}
        \begin{align}
            \bm{u}|_{r = a} &= \bm{U} + \bm{v}_A,\\
            \hat{\bm{u}}|_{r = a} &= \hat{\bm{U}} + \hat{\bm{v}},
        \end{align}
    \end{subequations}
    where $\bm{v}_A$ and $\hat{\bm{v}}$ represent the slip velocities for both problems.
    
    Since microswimmers move under force-free and torque-free conditions, \Eqref{eq:theorem} leads to 
    \begin{align}\label{eq:theorem_speed}
        \hat{\bm{F}}\cdot\bm{U} = -\int_{S}\bm{n}\cdot\hat{\bm{\sigma}}\cdot\bm{v}_A\ dS + \int_{S}\bm{n}\cdot\bm{\sigma}\cdot\hat{\bm{v}}\ dS
        +\int_{V}\bm{\Sigma}:\bm{\nabla}\hat{\bm{u}}\ dV,
    \end{align}
    where 
    \begin{align}
        \hat{\bm{F}} = \int_S \bm{n}\cdot\hat{\bm{\sigma}}\ dS   
    \end{align}
    is the hydrodynamic force acting on a passive particle with velocity $\hat{\bm{U}}$ in the Stokes fluid.
    Due to the linearity of the Stokes equations, we can express 
    \begin{align}
        \hat{\bm{u}} = \hat{\bm{L}}\cdot\hat{\bm{U}},\qquad
        \hat{\bm{\sigma}} = \hat{\bm{T}}\cdot\hat{\bm{U}},\qquad
        \hat{\bm{F}} = -\hat{\bm{R}}\cdot\hat{\bm{U}},
    \end{align}
    where the resistance tensor $\hat{\bm{R}}$ can be represented as
    \begin{align}
        \hat{\bm{R}} = -\int_{S}\bm{n}\cdot\hat{\bm{T}}\ dS.
    \end{align}
    Utilizing these linear relationships, \Eqref{eq:theorem_speed} yields swimming speeds given by
    \begin{align}\label{eq:theorem_vel}
        \begin{split}
        -\hat{\bm{R}}\cdot\hat{\bm{U}}\cdot\bm{U} =
        &-\int_{S}\bm{n}\cdot(\hat{\bm{T}}\cdot\hat{\bm{U}})\cdot\bm{v}_A\ dS + \int_{S}\bm{n}\cdot\bm{\sigma}\cdot\hat{\bm{v}}\ dS\\
        &+\int_{V} \bm{\Sigma}:\bm{\nabla}(\hat{\bm{L}}\cdot\hat{\bm{U}})\ dV.
        \end{split}
    \end{align}

    We consider two types of microswimmers with different swimming mechanisms: (i) a slip-driven squirmer~\cite{Lighthill1952-ui, Blake1971-ws} and (ii) a traction-driven squirmer~\cite{Daddi-Moussa-Ider2023-tv, Hosaka2023-tt}.
    \subsection{Slip-driven Squirmer}
        To solve the problem of a slip-driven squirmer with a prescribed surface velocity $\bm{v}_A$, we choose a spherical particle with a no-slip boundary condition $\hat{\bm{v}}_{\rm NS} = 0$ as an auxiliary problem~\cite{Stone1996-mx}. This choice eliminates the second term on the right-hand side of~\Eqref{eq:theorem_vel}, leading to the swimming speed, represented as
        \begin{align}
            \bm{U} = \hat{\bm{R}}^{-1}_{\rm NS}\cdot\left[\int_{S}\bm{n}\cdot\hat{\bm{T}}_{\rm NS}\cdot\bm{v}_A\ dS - \int_{V} \bm{\Sigma}:\bm{\nabla}\hat{\bm{L}}_{\rm NS}\ dV\right].
        \end{align}
        For a no-slip particle, $\hat{\bm{L}}_{NS}$, $\hat{\bm{T}}_{NS}$ and $\hat{\bm{R}}_{NS}$ are given by (see Appendix~\ref{appendix:no-slip} for a derivation),
        \begin{subequations}
            \begin{gather}
                \hat{\bm{L}}_{\rm NS} = \frac{3a}{4}\left(\frac{\bm{1}}{r}+ \frac{\bm{nn}}{r}\right) + \frac{a^3}{4}\left(\frac{\bm{1}}{r^3} - \frac{3\bm{nn}}{r^3}\right),\\
                \bm{n}\cdot\hat{\bm{T}}_{\rm NS}|_{r = a} = -\frac{3\eta}{2a}\bm{1}\\
                \hat{\bm{R}}_{\rm NS} = 6\pi\eta a\bm{1}.
            \end{gather}
        \end{subequations}
        We then obtain the swimming speed as follows:
        \begin{align}\label{eq:slip-driven}
            \bm{U} = -\frac{1}{4\pi a^2}\int_{S}\bm{v}_A\ dS - \frac{1}{6\pi\eta_0 a}\int_{V}\bm{\Sigma}:\bm{\nabla}\hat{\bm{L}}_{\rm NS}\ dV.
        \end{align}
        The first term represents the swimming velocity solution in the Stokes limit~\cite{Stone1996-mx}. The second term accounts for the non-Newtonian effect, which arises from the shear-rate-dependent viscosity.
        
\begin{figure}[tb!]
            \centering
            \includegraphics[width=\linewidth]{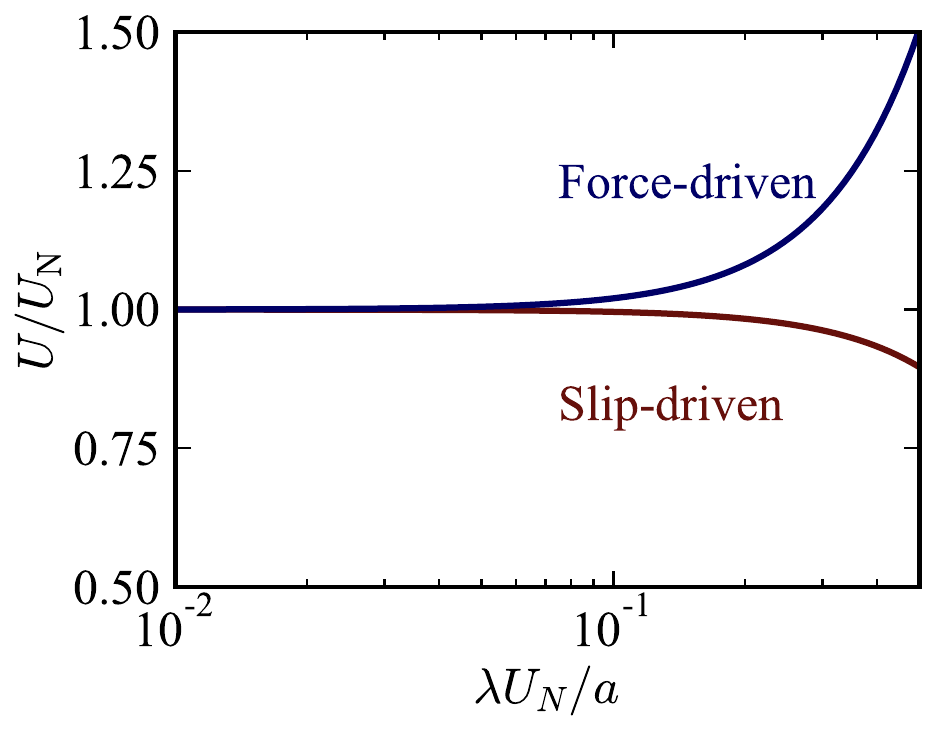}
            \caption{Swimming speeds of two types of microswimmers: a slip-driven squirmer and traction-driven squirmer. In shear-thinning fluids, traction-driven squirmers move faster, whereas slip-driven squirmers move slower than they do in Newtonian fluids. The viscosity ratio is $\beta = \eta_s / \eta_0 = 0$, and the power law index in the Carreau--Yasuda model is $n = 0.25$.
            }
            \label{fig:vel}
        \end{figure}

        Here, we consider a microswimmer with a surface slip velocity, denoted as $\bm{v}_A = -v_0 (\bm{1} - \bm{nn})\cdot\mathbf{p}$, where $\mathbf{p}$ is the swimming direction. The swimming speed of a slip-driven squirmer is given by~\cite{Datt2015-yc}.
        \begin{align}\label{eq:speed1}
            \bm{U} = \bm{U}_N - \frac{36}{65}\left(\frac{\lambda U_N}{a}\right)^2(1 - \beta)(1 - n)\bm{U}_N,
        \end{align}
        where $\bm{U}_N = U_N\mathbf{p} = (2 / 3)v_0\mathbf{p}$ is the swimming speed of a slip-driven squirmer in the Stokes fluid, and $\beta = \eta_\infty / \eta_0$ is the viscosity ratio. 
        For a shear-thinning fluid ($n < 1$), the swimming speed is slower than the Newtonian swimming speed $\bm{U}_N$ (see~\Figref{fig:vel}) and decreases as the viscosity ratio $\beta$ increases.
        In contrast, for a shear-thickening fluid ($n > 1)$, the swimming speed is faster than the Newtonian swimming speed $\bm{U}_N$.
        The swimming speed also depends on the swimming type (pusher, neutral, or puller), but pushers and pullers have the same swimming speed~\cite{Datt2015-yc}. 
        
    \subsection{Traction-driven Squirmer}
        To solve the swimming problem of a traction-driven squirmer with a prescribed active force density $\bm{f}_A$, we choose a spherical particle with a perfect-slip boundary condition $\hat{\bm{f}}^\parallel_{\rm PS} = (\bm{1} - \bm{nn})\cdot\hat{\bm{\sigma}}_{\rm PS}\cdot\bm{n} = \bm{0}$ as an auxiliary problem~\cite{Nasouri2021-tf, Daddi-Moussa-Ider2023-tv, Hosaka2023-tt}. 
        For a perfect-slip particle, $\hat{\bm{L}}_{\rm PS}$, $\hat{\bm{T}}_{\rm PS}$ and $\hat{\bm{R}}_{\rm PS}$ are given by (see Appendix~\ref{appendix:perfect}).
        \begin{subequations}
            \begin{align}
            \hat{\bm{L}}_{\rm PS} &= \frac{a}{2r}(\bm{1} + \bm{nn}),\\
            \bm{n}\cdot\hat{\bm{T}}_{\rm PS}|_{r = a} &= -\frac{3\eta}{a}\bm{nn},\\
            \hat{\bm{R}}_{\rm PS} &= 4\pi\eta a\bm{1},
        \end{align}
        \end{subequations}
        and the slip velocity of a perfect-slip particle is
        \begin{align}
            \hat{\bm{v}} = \frac{1}{2}(\bm{nn} - \bm{1})\cdot\hat{\bm{U}}.
        \end{align}
        We consider an axisymmetric surface-driven microswimmer, resulting in $\bm{v}_A$ having only a tangential component and $\bm{n}\cdot\hat{\bm{T}}_{\rm PS}\cdot\bm{v}_A = 0$. \Eqref{eq:theorem_vel} leads to
        \begin{align}\label{eq:force-driven}
            \bm{U} = \frac{1}{8\pi\eta_0a}\int_{S}\bm{f}_A^\parallel\ dS -\frac{1}{4\pi\eta_0a} \int_{V}\bm{\Sigma}:\bm{\nabla}\hat{\bm{L}}_{\rm PS}\ dV,
        \end{align}
        
        We derive the flow field created by a traction-driven squirmer in Newtonian fluids governed by the Stokes equations (at zeroth order ${O}({\rm Cu}^0)$). At the zeroth order ${O}(\rm Cu^0)$, the non-Newtonian contribution is $\bm{\Sigma} = \bm{0}$.
        Here, we consider a prescribed force density, denoted as $\bm{f}_A^\parallel = f_0(\bm{1} - \bm{nn})\cdot\mathbf{p}$. The swimming speed of the squirmer is 
        \begin{align}
            \bm{U}_N = \int_{S}\bm{f}_A^\parallel\ dS=  \frac{f_0 a}{3\eta_0}\mathbf{p}.
        \end{align}
        
        For the velocity field around a traction-driven squirmer, we use a superposition of a Stokeslet and a source dipole and impose the boundary conditions of zero radial velocity at the surface $\bm{n}\cdot\bm{v}_A = 0$ and the prescribed force given by $\bm{f}^\parallel_A = f_0(\bm{1} - \bm{nn})\cdot\mathbf{p}$. According to the force-free condition, the velocity flow in the lab frame can be expressed solely by a source dipole, 
        \begin{align}
            \bm{u}_0 = \frac{a^3}{r^3}\bm{U}_N - \frac{3a^3}{2r^3}(\bm{1} - \bm{nn})\cdot\bm{U}_N,
        \end{align}
        which is equivalent to the flow field created by a slip-driven squirmer in Stokes fluids.

        At first order ${O}({\rm Cu}^2)$, the swimming speed of a traction-driven squirmer is 
        \begin{align}
            \bm{U} = \bm{U}_N +\frac{27}{10}\left(\frac{\lambda U_N}{a}\right)^2(1 - \beta)(1 - n)\bm{U}_N.
        \end{align}
        A traction-driven squirmer moves slower/faster in shear-thickening/thinning fluids than in Newtonian fluids (see~\Figref{fig:vel}). In contrast, a slip-driven squirmer moves faster/slower in shear-thickening/thinning fluids (see~\Eqref{eq:speed1}).
        To understand why two types of squirmers exhibit opposite trends in fluids with shear-rate-dependent viscosities, we can decompose the swimming problem into two parts: drag and thrust~\cite{Datt2015-yc}. For a slip-driven squirmer, the increase/reduction in thrust caused by the shear-thickening/thinning property is greater than the increase/reduction in drag. This makes the squirmer faster/slower in shear-thickening/thinning fluids than in Newtonian fluids~\cite{Datt2015-yc}. 
        In contrast, the thrust force remains constant for a traction-driven squirmer in both shear-thickening/thinning and Newtonian fluids. The drag force, on the other hand, increases/decreases in shear-thickening/thinning fluids. As a result, a traction-driven squirmer can swim slower/faster in shear-thickening/thinning fluids than in Newtonian fluids.
        
\section{Breakdown of the scallop theorem}
    \begin{figure*}[tb!]
        \centering
        \includegraphics[width=0.8\linewidth]{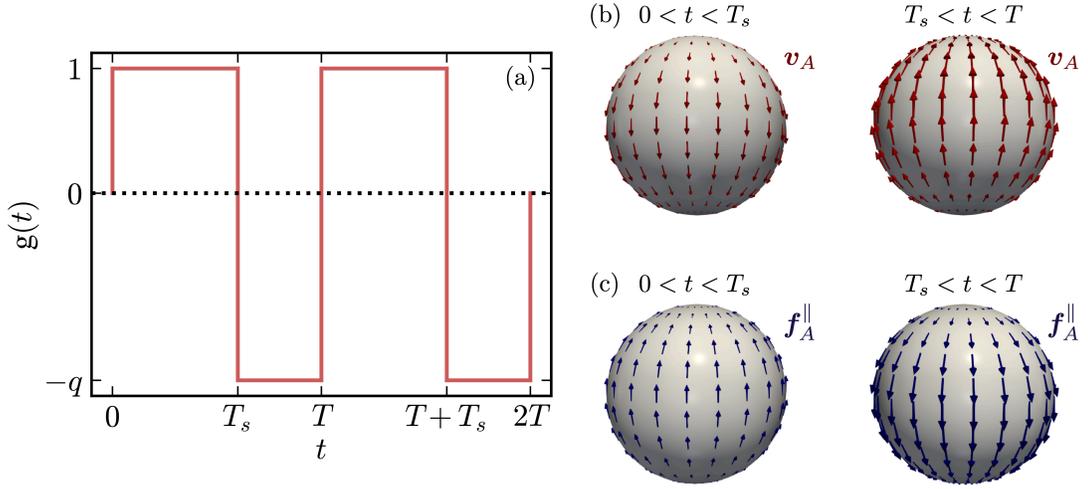}
        \caption{(a) The function $g(t)$, which describes the time-reversible deformation, as given in~\Eqref{eq:g}. Schematic representations of (b) a slip-driven squirmer with time-reversible deformation and (c) a traction-driven squirmer with time-reversible deformation. }
        \label{fig4}
    \end{figure*}
    In this section, we examine a time-reversible deformation and show that Purcell's scallop theorem~\cite{Purcell1977-ab} breaks down into two types of squirmers in fluids with shear-rate-dependent viscosity: (i) a slip-driven squirmer and (ii) a traction-driven squirmer.
    \subsection{Slip-driven Squirmer}
        For a time-reversible deformation, we consider a slip-driven squirmer with a slip velocity given by
        \begin{align}\label{eq:tr_slip}
            \bm{v}_A = -v_0(\bm{1} - \bm{nn})\cdot\mathbf{p}g(t),
        \end{align}
        where $g(t)$ is defined such that its average over one period $T$ is zero,
        \begin{align}
            \langle g(t) \rangle = \frac{1}{T}\int_0^Tg(t)\ dt = 0,
        \end{align}
        which reflects that the time-averaged slip velocity is zero $\langle\bm{v}_A\rangle = \bm{0}$. 
        Given \Eqref{eq:slip-driven}, the time-averaged swimming speed over one period $T$ is 
        \begin{align}
            \langle \bm{U} \rangle = - \frac{1}{6\pi\eta_0 a}\int_{V}\langle \bm{\Sigma}\rangle :\bm{\nabla}\hat{\bm{L}}_{\rm NS}\ dV.
        \end{align}
        At the first order ${O}({\rm Cu}^2)$, the time-averaged non-Newtonian contribution $\langle\bm{\Sigma}\rangle$ is
        \begin{align}
            \begin{split}
                \langle \bm{\Sigma} \rangle &= \frac{(\eta_0 - \eta_\infty)(n - 1)}{2}\lambda^2\langle |\dot{\gamma}_0|^2 \dot{\bm{\gamma}}_0\rangle\\
                &= \frac{(\eta_0 - \eta_\infty)(n - 1)}{2}\lambda^2|\dot{\gamma}'|^2\dot{\bm{\gamma}}'\langle g(t)^3\rangle.
            \end{split}
        \end{align}
        Since fluid inertia is negligible, the zeroth-order flow field $\bm{u}_0$ is
        \begin{align}
            \bm{u}_0 = \underbrace{\left[\frac{a^3}{r^3}\bm{U}_N - \frac{3a^3}{2r^3}(\bm{1} - \bm{nn})\cdot\bm{U}_N\right]}_{\bm{u}'}g(t),
        \end{align}
        and the corresponding strain-rate tensor $\dot{\bm{\gamma}}_0$ is 
        \begin{align}
            \dot{\bm{\gamma}}_0 = \bm{\nabla}\bm{u}_0 + (\bm{\nabla}\bm{u}_0)^T
            = \underbrace{\left[\bm{\nabla}\bm{u}' + (\bm{\nabla}\bm{u}')^T\right]}_{\dot{\bm{\gamma}}'}g(t),
        \end{align}
        where $\dot{\bm{\gamma}}'$ is the strain rate tensor corresponding to the flow velocity $\bm{u}'$. The time-averaged swimming speed of a slip-driven squirmer with the slip velocity described by \Eqref{eq:tr_slip} is
        \begin{align}
            \langle \bm{U} \rangle = -\frac{36}{65}\langle g(t)^3\rangle\left(\frac{\lambda U_N}{a}\right)^2(1 - \beta)(1 - n)\bm{U}_N
        \end{align}
        
        If $g(t)$ is a simple sinusoidal function, such as $g(t) = \sin\omega t$ or $\cos\omega t$, then $\langle g(t)^3\rangle = 0$, leading to no net locomotion in fluids with shear-rate-dependent viscosity. However, for a slip-driven squirmer in Oldroyd-B fluids, which exhibit viscoelasticity, the scallop theorem breaks down, resulting in net locomotion~\cite{Lauga2009-df}. Lauga noted that to achieve net movement with sinusoidal motion, the slip velocity distribution must be asymmetrical from front to back~\cite{Lauga2009-df}.

        Next, inspired by an experimental study by Qiu et al.~\cite{Qiu2014-nm}, we use two different slip velocities. They designed a swimmer similar to a Purcell scallop that could move in both shear-thickening and shear-thinning fluids.
        The function $g(t)$ can be described as (see~\Figref{fig4}),
        \begin{align}\label{eq:g}
            g(t) = 
            \begin{cases}
                   1 & (0 < t < T_s)\\
                -q & (T_s < t < T)
            \end{cases},
        \end{align}
        where $q$ denotes the ratio between the slip velocities from $0$ to $T_s$ and from $T_s$ to $T$. 
        One full cycle $T$ consists of two parts: $T = T_s + T_f$, where $T_\alpha = 2\pi / \omega_\alpha$ $(\alpha\in[s, f])$, with $\omega_s = B_1 / a$ and $\omega_f = qB_1 / a$.
        Under these conditions, the time-averaged deviation from Newtonian behavior is
        \begin{align}
            \langle\bm{\Sigma}\rangle = q(1 - q)\frac{(\eta_0 - \eta_\infty)(n - 1)}{2}\lambda^2|\dot{\gamma}'|^2\dot{\bm{\gamma}}',
        \end{align}
and the time-averaged swimming speed $\langle U \rangle$ of the swimmer up to ${O}(Cu^2)$ is
        \begin{align}\label{eq:slip_ave}
            \langle \bm{U} \rangle = -\frac{36}{65}q(1 - q)\left(\frac{\lambda U_N}{a}\right)^2(1 - \beta)(1 - n)\bm{U}_N.
        \end{align}
        \Eqref{eq:slip_ave} explicitly shows the breakdown of the scallop theorem for a slip-driven squirmer with different surface slip velocities ($q \ne 1$) in a Carreau--Yasuda fluid. In the Newtonian limit where $\lambda = 0$ or $\beta = 1$, we have $\langle \bm{U} \rangle = \bm{0}$, which is consistent with the conditions under which the scallop theorem holds.
        
    \subsection{Traction-driven Squirmer}
        Next, we examine a traction-driven squirmer that moves with a time-reversible active force density, given by
        \begin{align}\label{eq:force_bc}
            \bm{f}_A^\parallel = f_0 (\bm{1} - \bm{nn})\cdot\mathbf{p} g(t).
        \end{align}
        Using \Eqref{eq:force-driven}, the time-averaged swimming speed of a traction-driven squirmer is
        \begin{align}
            \langle\bm{U}\rangle &= \frac{1}{8\pi\eta_0a}\int_{S}\langle\bm{f}_A^\parallel\rangle\ dS -\frac{1}{4\pi\eta_0a} \int_{V}\langle\bm{\Sigma}\rangle:\bm{\nabla}\hat{\bm{L}}_{\rm PS}\ dV\\ 
            &=-\frac{1}{4\pi\eta_0a} \int_{V}\langle\bm{\Sigma}\rangle:\bm{\nabla}\hat{\bm{L}}_{\rm PS}\ dV.
        \end{align}
        Similar to a slip-driven squirmer, if $g(t)$ is a simple sinusoidal function, then the net locomotion is zero. Therefore, we consider a traction-driven squirmer with two distinct active force densities expressed by \Eqsref{eq:g} and \eqref{eq:force_bc} (see~\Figref{fig4}).
        The time-averaged swimming velocity is
        \begin{align}
            \langle\bm{U}\rangle = \frac{27}{10}q(1 - q)\left(\frac{\lambda U_N}{a}\right)^2(1 - \beta)(1 - n)\bm{U}_N.
        \end{align}
        This demonstrates that Purcell's scallop theorem for a traction-driven squirmer breaks down in fluids governed by the Carreau--Yasuda constitutive equations. In the Newtonian limit ($\lambda = 0$ or $\beta = 1$), $\langle \bm{U} \rangle = \bm{0}$. 
        
        When $q > 1$, slip-driven squirmers can move backward/forward, whereas traction-driven squirmers can move forward/backward in shear-thickening/thinning fluids. This occurs because the traveling speed ($0 < t < T_s$) is slower/faster for slip-driven squirmers and faster/slower for traction-driven squirmers than their returning speed ($T_s < t < T$) in shear-thickening/thinning fluids.
        The absolute value of the time-averaged swimming speed of a traction-driven squirmer is greater than that of a slip-driven squirmer compared with the Newtonian swimming speed $\bm{U}_N$. This suggests that a traction-driven squirmer with time-reversible deformation can swim more easily in fluids governed by the Carreau--Yasuda constitutive equations.

\section{Conclusion}\label{sec:conclusion}
    In conclusion, we focused on microswimming problems via the reciprocal theorem and perturbation analysis and derived the speed of a traction-driven squirmer in Carreau--Yasuda fluids. Slip-driven squirmers move slower in shear-thinning fluids than in Newtonian fluids~\cite{Datt2015-yc}. However, we found that swimmers driven by active force densities move faster in shear-thinning fluids than in Newtonian fluids.
    Additionally, we demonstrated that Purcell's scallop theorem breaks down for both surface-driven microswimmers in shear-thickening and shear-thinning fluids. Inspired by previous experimental investigations~\cite{Qiu2014-nm}, we use two different slip speeds or force densities. The reciprocal swimming mechanism of the microscallop described herein may offer a universal approach for microswimming in biological fluids.
    For future work, our study will explore the collective behavior of surface-driven microswimmers in shear-thickening and shear-thinning fluids. We anticipate that the differences between slip-driven and traction-driven squirmers, as well as the various swimming types, will significantly impact the collective behaviors of squirmer suspensions in shear-thickening and shear-thinning fluids.

\begin{acknowledgments}
The authors would like to thank Raymond E. Goldstein and Ronojoy Adhikari for their fruitful discussions.
This work was supported by Grants-in-Aid for Scientific Research (JSPS KAKENHI) under Grant Number 20H05619, by the JSPS Core-to-Core Program ``Advanced core-to-core network for the physics of self-organizing active matter (JPJSCCA20230002)'' and by JST SPRING, Grant Number JPMJSP2110.
\end{acknowledgments}

\appendix
\section{Derivation of flow fields for a passive particle}
    We calculate the flow fields around a passive particle to determine the active swimming velocity of a microswimmer. We now consider the flow around a particle moving with velocity $\hat{\bm{U}}$. In the lab frame, the boundary conditions for a passive particle are given by
    \begin{subequations}\label{appendix:bc}
        \begin{align}
            &\hat{\bm{u}} = \bm{0}\quad(r \to \infty),\\
            &\bm{n}\cdot\hat{\bm{u}}|_{r = a} = 0.
        \end{align}
    \end{subequations}
    
\subsection{No-slip particle}\label{appendix:no-slip}
    We revisit the flow field around a no-slip particle~\cite{Happel1983-ko}, which serves as an auxiliary problem to derive the swimming speed of a slip-driven squirmer.
    In the lab frame, the boundary condition for a no-slip particle is
    \begin{align}\label{appendix:bc1}
        \hat{\bm{u}}|_{r = a} = \hat{\bm{U}}.
    \end{align}
    The Stokeslet $\bm{G}$ and source dipole $\bm{D}$ and the corresponding stress tensors $\bm{T}^G$ and $\bm{T}^D$ are given by~\cite{Pozrikidis1992-lu},
    \begin{subequations}\label{appendix:GD}
        \begin{align}
            \bm{G} &= \frac{\bm{1}}{r} + \frac{\bm{nn}}{r},\\
            \bm{D} &= -\frac{1}{2}\bm{\nabla}^2\bm{G} =  -\frac{\bm{1}}{r^3} + 3\frac{\bm{nn}}{r^3},\\
            T_{ijk}^G &= -6\eta \frac{n_in_jn_k}{r^2},\\
            T_{ijk}^D &= 6\eta\frac{\delta_{ij}n_k + \delta_{ik}n_j + \delta_{jk}n_i}{r^4} - 30\eta\frac{n_in_jn_k}{r^4}.
        \end{align}
    \end{subequations}
    For the velocity and stress tensor around a perfect-slip particle, we assume a combination of a Stokeslet and a source dipole:
    \begin{align}
        \hat{u}_i &= G_{ij}A_j + D_{ij}B_j,\\
        \hat{\sigma}_{ik} &= T_{ijk}^G A_j + T_{ijk}^D B_j.
    \end{align}
    By applying the boundary conditions (\Eqsref{appendix:bc} and~\eqref{appendix:bc1}), we find that $\bm{A} = 3a\hat{\bm{U}} / 4$, $\bm{B} = -a^3\hat{\bm{U}} / 4$. The flow velocity and stress tensor are then given by 
    \begin{subequations}
        \begin{align}
            \hat{\bm{u}} &= \left[\frac{3a}{4}\left(\frac{\bm{1}}{r}+ \frac{\bm{nn}}{r}\right) + \frac{a^3}{4}\left(\frac{\bm{1}}{r^3} - \frac{3\bm{nn}}{r^3}\right)\right]\cdot\hat{\bm{U}}\\
            \hat{\bm{\sigma}} &= \left(\frac{3a}{4}\bm{T}^G -\frac{a^3}{4}\bm{T}^D\right)\cdot\hat{\bm{U}}.
        \end{align}
    \end{subequations}
    The traction on the surface and the total force acting on the particle are obtained:
    \begin{subequations}
        \begin{align}
            \hat{\bm{f}} &= \hat{\bm{\sigma}}|_{r = a}\cdot\bm{n} = -\frac{3\eta}{2a}\hat{\bm{U}},\\
            \hat{\bm{F}} &= \int_{S}dS\ \hat{\bm{f}} = -6\pi\eta a \hat{\bm{U}}.
        \end{align}
    \end{subequations}

\subsection{Perfect-slip particle}\label{appendix:perfect}
    Next, we consider the flow field around a perfect-slip particle, which is the auxiliary problem for deriving the swimming speed of a traction-driven microswimmer~\cite{Daddi-Moussa-Ider2023-tv}.
    In the lab frame, the boundary condition for a perfect-slip particle is
    \begin{align}\label{appendix:bc2}
        (\bm{1} - \bm{nn})\cdot\hat{\bm{\sigma}}|_{r = a}\cdot\bm{n} = \bm{0}.
    \end{align}
    For the velocity and stress tensor around a perfect-slip particle, we again assume a superposition of a Stokeslet and a source dipole,
    \begin{align}
        \hat{u}_i &= G_{ij}A_j + D_{ij}B_j,\\
        \hat{\sigma}_{ik} &= T_{ijk}^G A_j + T_{ijk}^D B_j.
    \end{align}
    Imposing boundary conditions (\Eqsref{appendix:bc} and~\eqref{appendix:bc2}), we find $\bm{A} = a\hat{\bm{U}} / 2$, $\bm{B} = \bm{0}$. The flow velocity and stress tensor are then given by 
    \begin{align}
        \hat{\bm{u}} =  \frac{a}{2r}\left(\bm{1} + \bm{nn}\right)\cdot\hat{\bm{U}},\qquad
        \hat{\bm{\sigma}} = -3\eta a(\bm{n}\cdot\hat{\bm{U}})\frac{\bm{nn}}{r^2}.
    \end{align}
    The flow velocity on the particle surface is 
    \begin{align}
        \hat{\bm{u}}|_{r = a} = \hat{\bm{v}} = \frac{1}{2}(\bm{nn} - \bm{1})\cdot\hat{\bm{U}},
    \end{align}
    and the traction at the surface and the total force acting on the particle are given by
    \begin{subequations}
        \begin{align}
            \hat{\bm{f}} &= \hat{\bm{\sigma}}|_{r = a}\cdot\bm{n} = -\frac{3 \eta}{a}(\bm{n}\cdot\hat{\bm{U}})\bm{n},\\
            \hat{\bm{F}} &= \int_{S}dS\ \hat{\bm{f}} = -4\pi\eta a \hat{\bm{U}}.
        \end{align}
    \end{subequations}

\bibliography{ref.bib}

\end{document}